\documentstyle[11pt,newpasp,twoside,epsf]{article} 
\def\edcomment#1{\iffalse\marginpar{\raggedright\sl#1\/}\else\relax\fi}
\reversemarginpar

\newcommand{\be}{\begin{equation}}
\newcommand{\ee}{\end{equation}}
\newcommand{\bea}{\begin{eqnarray}}
\newcommand{\eea}{\end{eqnarray}}

\begin{document}
\title{The Formation and Evolution of Multiple Star Systems}
 \author{Sverre J. Aarseth}
\affil{Institute of Astronomy, University of Cambridge, Cambridge, UK}
\author{Rosemary A. Mardling}
\affil{Mathematics Department, Monash University, Melbourne, Australia}

\begin{abstract}
Multiple systems play an important role in the evolution of star clusters.
First we discuss several formation mechanisms which depend on the presence
of binaries, either primordial or of dynamical origin.
Hierarchical configurations are often stable over long times and yet may
experience evolution of the internal orbital parameters.
We describe an attempt to model the eccentricity change induced by the
outer component using an averaging method, together with the effects due
to tidal dissipation and apsidal motion acting on the inner binary.
This treatment is adopted for systems with high induced eccentricity which
gives rise to some interesting outcomes of significant period shrinkage.
\end{abstract}

\section{Introduction}

$N$-body simulations of star clusters need to include a significant
fraction of primordial binaries in order to be realistic
(Raboud \& Mermilliod 1999).
This introduces a new complexity into the problem and also poses many
technical challenges.
Among the main computational aspects we mention strong interactions and
short time-scales, as well as astrophysical phenomena involving tidal
interactions or mass transfer in short-period binaries.
In the present paper, we are concerned with the formation and evolution
of multiple systems containing different hierarchical configurations.
Let us introduce the notation B and S for binaries and single stars.
In addition to standard triples, denoted by [B,S], we distinguish quadruples
in the form [B,B] as well as higher-order systems, such as [[B,S],S] up to
sextuplets [[B,B],B] which are all recognized by the algorithms.

It is known from scattering experiments that hierarchical triples may form
as the result of binary-binary collisions, where the fourth component is
ejected (Mikkola 1983).
Provided the pericentre of the outer orbit is sufficiently wide, such
configurations may be stable over long times (Mardling \& Aarseth 1999).
Hence it only requires a small formation rate in order to build up a
substantial population of such systems.
Given long life-times, more complex structures may be formed as the result
of favourable interactions.
In the present contribution, we consider the formation process in more detail
by examining some numerical results and also model the inner binary
evolution which is subject to eccentricity modulations.

\section{Star Cluster Modelling}

The modelling of star clusters involves many ingredients.
Here we summarize some of the most relevant aspects.
Since rich open clusters are studied at present, we adopt an initial
half-mass radius $r_h \simeq 3$~pc for $N_s = 8000$ single stars.
In addition, $N_b = 2000$ primordial binaries are included in the standard
models, although smaller fractions were used earlier.
A linearized tidal field for circular orbits is employed in the equations
of motion, with a tidal radius of about $30~pc$.
The semi-major axes $a_i$ are chosen from a flat distribution in $\log a$
(Kroupa 1998) with lower period limit of $4~d$.
For the upper limit we take $a_{\rm max} \simeq 240~AU$ which exceeds the
semi-major axis of a typical hard binary, $a_{\rm hard} \simeq 2 r_h/N$, by
a factor of 2.
Finally, the single stars and binary component masses are sampled from an
IMF based on the solar neighbourhood in the range $[15, 0.2]~m_{\sun}$
(Kroupa, Tout, \& Gilmore 1993).
This produces an initial cluster model with $N = 12,000$ stars, total mass
$8100~m_{\odot}$ and crossing time of $3.7~Myr$, where the latter is
defined from twice the virial radius divided by the rms velocity.

On the astrophysical side, we employ synthetic stellar evolution to describe
the tracks in the HR diagram (Tout et al. 1997).
In particular, changes of stellar radius and mass are modelled for
solar-type abundances.
Finite-size effects due to tidal circularization and Roche-lobe mass transfer
are also taken into account
(Mardling \& Aarseth 2001, Tout et al.\ 1997).
Finally, stellar spins have recently been included for hard binaries during
the circularization stage.
This implementation is based on Hut (1981) theory which assumes that the
stellar rotation axes are perpendicular to the orbital plane.
Since the total angular momentum is constant, we obtain the new semi-major
axis by solving an equation for the eccentricity which contains the spins,
where the latter are integrated separately.

The actual equations of motion for the stars are advanced by several
techniques.
Here it is relevant to note that hard binaries are studied by the classical
Kustaanheimo-Stiefel (1965, hereafter KS) regularization method which has
been adapted to an accurate Taylor series formulation
(Mikkola \& Aarseth 1998).
Weakly perturbed binaries can now be integrated over long times without
any significant secular errors.
Strong interactions between binaries and single stars or other binaries are
treated by the powerful chain regularization (Mikkola \& Aarseth 1993).
This method is particularly useful for the present work concerned with
identifying hierarchies.
In addition, binaries of arbitrarily small period are handled by the
so-called slow-down technique based on adiabatic invariants which entails
modifications of the chain regularization algorithm
(Mikkola \& Aarseth 1996).

The smooth workings of all the relevant numerical procedures requires a
major undertaking which is essentially completed.
It is particularly advantageous to perform the single particle integrations
on the HARP special-purpose computer (Aarseth 1996) since this allows
larger $N$-values to be reached.
In principle, any binary fraction can be studied, although high values are
relatively expensive because of the additional interactions on short
time-scales, as well as astrophysical complications which must be handled
on the host.
However, the adopted distribution does not include a significant proportion
of soft binaries which are likely to be present initially, and hence the
effective population is quite large.

The overall evolution of a cluster model is characterized by a core-halo
structure with a growing degree of mass segregation.
Because of mass loss from evolving stars, the core maintains relatively
low density such that the energy binding the cluster is significantly
diminished.
Contrary to expectations based on conservative models (McMillan \& Hut 1994),
the binary fraction now increases in the core (cf.\ Aarseth 1996) and even
the hierarchical systems tend to be centrally concentrated.
It is also a general feature of cluster dynamics that interactions involving
hard binaries often result in high-velocity escapers of single stars as well
as binaries.
Although there are technical complications of combining different schemes
efficiently, it is undoubtedly beneficial to employ regularization
methods for studying such energetic processes.
In addition, the description in terms of dominant two-body motions also
serves as a natural tool box for analyzing stable systems.
Hence using well behaved equations of motion increases the confidence in
the outcomes while at the same time facilitating the data analysis.

\section{Formation of Hierarchies}

Simulations  with primordial binaries frequently involve binary-binary
collisions with the subsequent formation of long-lived triples.
This is only one of several channels which may yield such outcomes but
for technical reasons, it is the easiest one to identify.
Thus if the binaries are sufficiently hard, such interactions are studied
by chain regularization and a check for hierarchical configurations is
carried out at every stage if certain distance ratios are satisfied.
The question then arises whether such systems might be stable on long
time-scales which would warrant a more efficient description.

Interest in the stability of hierarchical triples was originally motivated
by trying to understand the isolated three-body problem.
An early numerical investigation by Harrington (1972) produced a
semi-empirical fitting formula for stability, albeit with a restricted set
of parameters.
Generalization of this formula to different masses (Bailyn 1984) served a
useful purpose in $N$-body simulations for many years.
If a triple is defined to be stable, the inner binary is replaced
temporarily by its centre of mass so that the outer component can become
part of a new KS solution, thereby speeding up the calculation which might
otherwise be prohibitive.
In this approximation, we neglect any secular evolution of the inner binary
which is justified theoretically for stable systems.
The stability criterion was improved by Eggleton and Kiseleva (1995) who
considered a wider range of parameters but still limited to coplanar motion
and circular outer orbits.
The resulting fitting formula (corrected for a typographical error) has
been in use during recent years.
Most systems with small outer eccentricity accepted for the special KS
treatment usually satisfy the stability criterion by a wide margin.
However, a number of other border-line cases are invariably time-consuming
and it is therefore desirable to obtain sharper criteria.

A completely new way of looking at stability, based on the binary-tides
problem (Mardling 1995), has resulted in a semi-analytical criterion which
applies to a wide range of outer mass ratios and arbitrary outer
eccentricities (Mardling \& Aarseth 1999).
Here the boundary for the outer pericentre distance, $R_p^{\rm crit}$, is
given in terms of the inner semi-major axis, $a_{\rm in}$, by
\be
{R_p^{{\rm crit}}}
~=~ C\left[(1+q_{{\rm out}})\frac{(1+e_{{\rm out}})}
{(1-e_{{\rm out}})^{1/2}}\right]^{2/5} a_{{\rm in}} \,,
\ee
where $q_{\rm out} = m_3/(m_1 + m_2)$ is the outer mass ratio, $e_{\rm out}$
is the corresponding eccentricity and $C\simeq 2.8$ is determined
empirically.
This expression can now be understood from first principles (see Mardling
this volume).
Again this criterion is valid for coplanar prograde orbits
and ignores a weak dependence on the inner eccentricity and inner mass ratio.
Since inclined systems are more stable, we include a linear  correction
factor of up to 30~\% for retrograde motion, in qualitative agreement
with an earlier study (Harrington 1972) and recent experiments.

The case of a quadruple system (i.e.\ [B,B]) is handled in a similar way.
Here the outer binary represents the third body and, in analogy with the
binary tides problem for two extended objects, we include a small additional
correction factor to account for the size.
Likewise, higher-order systems of the type [[B,S],S], etc may form during
the post-collapse phase when the density in the inner region has decreased.
Although the binding energy of new KS binary must be significant in order to
be accepted, there is often ample hierarchical space for three or four
hierarchical levels unless the respective outer eccentricities are large.

The advantage of studying hierarchies by a more efficient method has been
recognized for a long time (cf.\ Aarseth 1985).
Up to now, only some main results of such models have been given
(Kiseleva et al.\ 1996, de la Fuente Marcos et al.\ 1997) without discussing
the dynamical aspects. 
Figure~1 shows the time evolution of a typical standard model defined above.
The early stage is characterized by an increase of stable systems which are
absent initially until a peak of about 16 members is reached, followed by a
gradual decline reflecting the decreasing particle number.

\begin{figure}
\plotfiddle{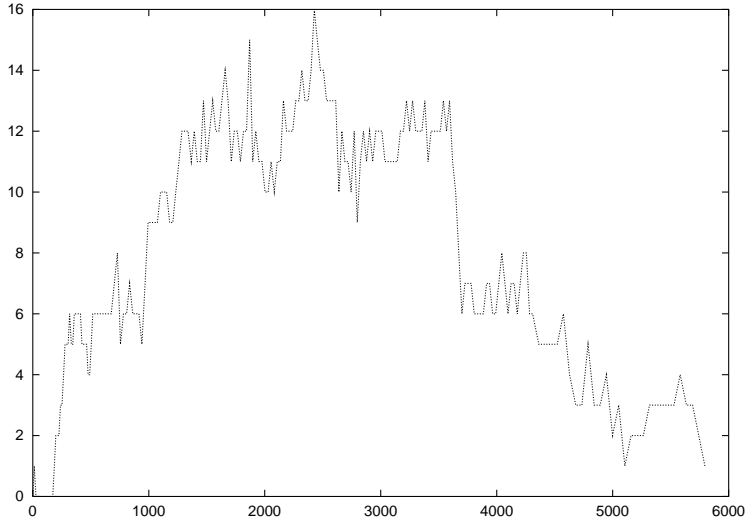}{6.5cm}{270}{40}{40}{-160}{220}
\caption{Evolution of stable hierarchical systems (time in Myr).}
\end{figure}

In Figure~2 we display another standard model together with one which has
half the membership; i.e.\ $N_s = 4000$ and $N_b = 1000$.
Again a significant population is established and the smaller system mirrors
the behaviour on a shorter time-scale.
In either case the total life-times of the cluster models are displayed;
i.e.\ the calculation terminates when $N < 10$.

\begin{figure}
\plotfiddle{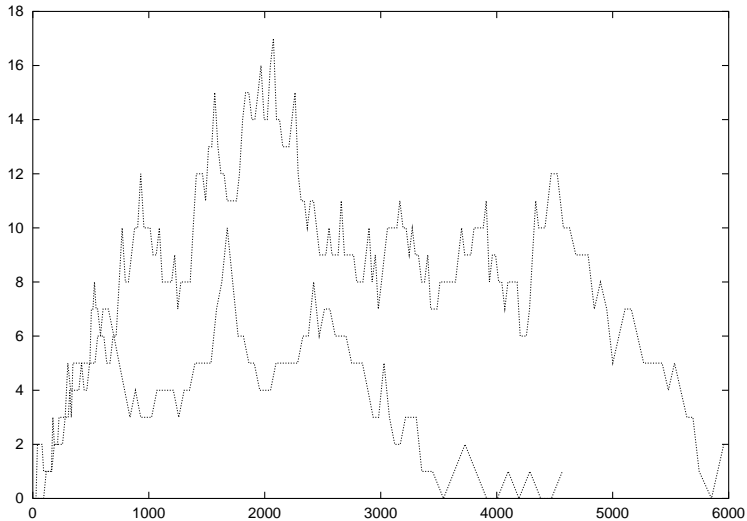}{6.5cm}{270}{40}{40}{-160}{220}
\caption{Stable hierarchical systems, $N$ = 12,000 and 6000.}
\end{figure}

A closer look at the data for the two standard models above reveals 300 and
288 new independent hierarchical systems.
Here we define a hierarchy as new if it has not been recorded during the
previous crossing time.
Moreover, the direct formation of 14 and 30 stable triples, respectively,
were identified as the end products of chain regularization.
This may be compared to about 630 chain regularization events in which
130 and 152 were associated with binary-binary interactions.
Although most of these stable triples originated in binary-binary collisions,
about one quarter were associated with chain interactions involving three
members.
However, such outcomes may still be attributed to binary-binary collisions
in which the widest binary ceases to be treated by the KS method immediately
before the chain regularization is activated.
Undoubtedly many more triples form in less extreme interactions where the
conditions of compact size used to initiate chain regularization do not
apply.

The process of triple formation has been examined further in order to test
the supposition above.
This involved re-calculating the late stages of a smaller model
($N_s = 4000, \, N_b = 1000$ at $t \simeq 1.5~Gyr$) with additional
diagnostics.
At each termination of a strongly perturbed KS solution due to another
binary we record the relevant particle identities.
It is then a simple matter to compare this data with the next stable triple.
Somewhat surprisingly, more than three quarters of about 80 terminations
are associated with new triples, although many are short-lived.
This high rate of temporary triple formation in hard binary-binary
interactions are in qualitative accord with earlier work.
Thus Mikkola (1983, 1984), excluding fly-by's, found the probability
increased from $20~\%$ to $50~\%$ when going from equal binding energies to
a ratio of 4, whereas Bacon, Sigurdsson, \& Davies (1996) reported about
$10~\%$ stable triples for their scattering experiments with a wide range
of impact parameters.

The formation of higher-order systems is of considerable interest.
Quadruples of the type [B,B] which satisfy Eq.~(1) are relatively common,
although many are of short duration.
Guided by examination of the model data presented above and based on
sampling intervals of $12~Myr$ at early and $24~Myr$ at later times, we
distinguish two different modes of formation.
Thus in some cases the initial binding energies of the relative centre of
mass motions tend to be near the imposed energy boundary of one quarter of
a typical hard binary.
Depending on the evolutionary state of the cluster, the characteristic
outer periods are $\ge 10^6~d$.
There is also some evidence to suggest a gradual shrinkage of the external
orbit for even larger periods until the stability conditions are met but
this process needs to be studied in more detail.
The second formation channel is more readily identified and involves the
exchange of a single binary with the outer component of a stable triple.
Conversely, an existing strongly bound quadruple can be disrupted by
exchange with either another binary or a single body.

The formation of quadruples in the two large models may be summarized as
follows.
The first model only revealed one significant system before epoch $2.2~Gyr$.
Subsequently, five energetic quadruples formed by exchange, whereas only a
few long-lived systems with relatively wide orbits were identified.
In fact, one of the exchange events involved the collision of two stable
triples, resulting in escape.
Here the relevant life-times were in the range $30 - 700 ~Myr$.
The second model contained one early quadruple lasting $200~Myr$.
Again the later stages ($>~2~Gyr$) exhibited some characteristic activity,
with about ten systems formed via exchange.
It should be noted that both binaries in quadruples often have short
periods so that their relative orbits are best described as two-body
systems.
Viewed in this way, the small proportion of such systems relative to
triples is connected with the low probability of new binaries being formed
in the presence of a primordial population (cf.\ Aarseth 1996).

Several other types of hierarchies have also been observed in the
simulations, particularly during the later stages.
Thus the two types of quintuplets [[B,B],S] and [[B,S],B], as well as
several sextuplets of the type [[B,S],[B,S]] with significant life-times
have been noted.
Finally, one example of the classical sextuplet [[B,B],B] has also been
identified in a recent cluster model.

The first significant evidence of a [[B,S],B] quintuplet in a model with
$N_s = 4000, \, N_b = 1000$ occurred at age $2.5~Gyr$ when $N_s \simeq 500$.
This system remained stable over some $3~Myr$ with a relatively short outer
period of about $6 \times 10^4~yr$.
Subsequently, at age $3.5~Gyr$ another system of this type appeared, also
containing the same circularized inner binary of period $1.1~d$.
This quintuplet survived during the next $\simeq 2 \times 10^8~yr$ until
complete disruption of the cluster, with sporadic terminations due to the
updating of the stellar evolution parameters.
Now the outer period was quite large ($3~Myr$) but the system avoided
disruption by being massive and staying outside the core.

Returning to the first quintuplet mentioned above, some light can be shed
on its pre-history leading up to the formation.
Following a binary-binary collision involving the small circularized binary,
a triple with $T_{\rm out} \simeq 1.2 \times 10^4~d$ was created by
exchange at epoch $1.8~Gyr$.
This strongly bound triple later interacted with a binary where one
component was exchanged, leading to $T_{\rm out} \simeq 4 \times 10^4~d$
which remained approximately constant until another exchange, giving
$T_{\rm out} \simeq 2 \times 10^4~d$.
Finally, the actual formation of the quintuplet can be traced to the triple
acquiring an outer component which itself was a binary.
Here the relative orbit of the two subsystems was a factor of 100 larger
than the size of the triple, which suggests a dynamical friction mode of
formation involving two massive bodies near the cluster centre
(cf.\ Aarseth 1972).

The deficiency of [[B,B],S] quintuplets as well as classical sextuplets in
the present preliminary investigation based on only a few models may be
understood by the following qualitative considerations.
Thus the predominant occurrence of the type [[B,S],B] appears to be
connected with the formation processes of the inner systems, since most
quadruples tend to be less compact than triples.
This implies that there is less hierarchical space for the next level of
the hierarchy to be stable, in qualitative agreement with our tentative
findings.
In addition, there is also a smaller population of quadruples acting as
seeds.

We end this section by highlighting the problem of long-lived hierarchies
which do not satisfy the standard stability criterion even after the
inclination effect has been included.
Thus in one recent example, the {\it outer} semi-major axis of a persistent
triple was $18~AU$, compared to the current hard binary size of $750~AU$.
Since the corresponding outer period was only $36~yr$, the calculation of
this configuration proved very time-consuming and produced significant
systematic errors due to direct integration of the third body.
Yet the outer pericentre was inside the value of $R_p^{{\rm crit}}$ by a
factor 0.45 so that the inclination modification of 0.75 was not sufficient
for stability to be accepted here.
Further analysis of the data revealed a semi-major ratio of 350 and a large
value of the outer eccentricity, with $e_{\rm out} = 0.987$.

In the case of high outer eccentricity and in analogy with tides in
binaries, the amount of energy exchanged during a pericentre passage of the
outer body is proportional to ($a_{\rm in} / R_p)^6$.
Hence the small energy exchange per passage resulting from a small ratio
would take a long time for the random walk process to reach
$e_{\rm out} > 1$.
In order to accommodate such systems, we introduce an experimental concept
of practical stability, with the number of outer periods given by
\be
{\cal N}_{\rm out} \,=\, 1 + 10 \,e_{\rm out} / (1 - e_{\rm out}) \,.
\ee
Implementing this for $e_{\rm out} > 0.96$ achieved the desirable result.
Thus the present example demonstrates the need for more sophisticated
decision-making to deal with unusual configurations which would otherwise
introduce spurious errors and also slow down the calculation unnecessarily.

The earlier history of the present example illustrates another interesting
aspect of hierarchical dynamics.
Thus the inner binary was part of a long-lived wide triple
($T_{\rm out} \simeq 1 \times 10^6~d$) before interacting with a much more
compact triple ($T_{\rm out} \simeq 5 \times 10^4~d$).
In analogy with binary-binary collisions leading to triples by exchange,
this produced a stable quintuplet after the second triple was exchanged
with the single outer component.
In fact, the new quintuplet survived during the next $2~Myr$ with an
outermost period of $\simeq 2 \times 10^6~d$.
Because of the large distance ratios, such a configuration is essentially
a more loosely bound triple which may become unstable by an increase of
the outer eccentricity.
Although still a hard binary by virtue of the large masses, external
perturbations resulted in termination, leaving behind the strongly bound
system discussed above.

\section{Averaging Treatment}

Left to itself, a stable triple system experiences changes of the outer
orbital elements due to external perturbations, whereas the internal binary
retains its characteristics.
However, in reality there is an exchange of angular momentum between the two
relative motions even if the inner semi-major axis is secularly constant.
Thus the outer body induces an eccentricity change of the inner orbit with a
corresponding response in the relative inclination.
This effect gives rise to so-called Kozai (1962) cycles which produce
significant changes above a certain value of the inclination.
From the constancy of angular momentum, together with $a = const$, the
relation between inclination and inner eccentricity is given by
\be
\cos^2 i \, (1 \,-\, e^2) \, = \, const \,.
\ee
Hence depending on the inclination immediately after the formation of a
hierarchical triple, there is a possibility of reaching a large value of
the inner eccentricity.

The corresponding period of the Kozai cycle is expressed in terms of the
respective orbital periods by (Heggie 1996)
\be
T_{\rm Kozai} \,=\, (1 \,-\,e_{{\rm out}}^2)^{3/2} \,
g(e_{{\rm in}}, i) \, T_{{\rm out}}^2/T_{{\rm in}} \,,
\ee
where $g$ is a function of order unity for the cases of interest here.
The maximum eccentricity, $e_{\rm max}$, may also be evaluated (Heggie 1996)
and plays an important part in the decision-making.
Here we concentrate on systems with large $e_{\rm max}$ which may become
tidally active following subsequent growth of the radii.
The evolution of such systems may then be modelled by combining the
relevant effects using averaging techniques.

The secular evolution of the inner binary of a hierarchical system is
calculated via the average of the rate of change of the Runge-Lenz vector
and the specific angular momentum vector.
The Runge-Lenz vector is given by
\be
{\bf e}_{\rm in} \,=\,
{\bf v} \times {\bf h} /(m_1+m_2) \,-\, {\bf r}/r \,,
\ee
where $\bf v$ is the relative velocity and ${\bf h} = {\bf r} \times {\bf v}$
is the specific angular momentum.
The rates of change of these vectors depend on the accelerations produced by
the tidal and spin bulges, tidal dissipation, a relativistic potential,
as well as the third body.
These effects are included to quadrupole order and the equations integrated
(averaged over one inner orbital period) are
\bea
<\dot {\bf e}> \,&=&\,
b_1 \, \widehat {\bf e} \,+\, 
b_2 \, \widehat {\bf q} \,+\, b_3 \, \widehat {\bf h} \\
<\dot {\bf h}> \,&=&\,
c_1 \, \widehat {\bf e} \,+\, 
c_2 \, \widehat {\bf q} \,+\, c_3 \, \widehat {\bf h} \,,
\eea
with $\widehat {\bf q} = \widehat {\bf e} \, \times \, \widehat {\bf h}$.
The contributions to $b_i$ and $c_i$ from the third body perturbation
were derived by Heggie (1996), while the tidal and spin contributions
are from Eggleton, Kiseleva, \& Hut (1998). 
Using units of solar radius and mass, the period of the relativistic
precession in years is
\be
T_{GR}\,=\,3.4\times 10^7 (1 - e^2)\,T_{in}\,a_{\rm in}/(m_1 + m_2) \,.
\ee
So far, the effect of stellar rotation has been ignored but this is
currently being rectified.
The new orbital elements are derivable from ${\bf e}$ and ${\bf h}$.
In this procedure we assume that the outer component is stationary during
each interval and the back reaction is not considered.
For the usual case of perturbed motion, the relevant time interval
corresponds to a small fraction of the outer period which results in an
appropriate sampling.

The different contributions depend sensitively on inverse powers of the
semi-major axis and eccentricity, with $b_1 \propto 1/a^8 (1 - e^2)^{13/2}$, 
$b_2 \propto 1/a^{13/2} (1 - e^2)^4$ from the tidal dissipation and
apsidal motion terms, respectively, whereas the relativistic precession
part of $b_2$ is $\propto 1/a^{5/2} (1 - e^2)$.
Consequently, the eccentricity growth due to the Kozai cycle may be detuned
by the apsidal motion or relativistic precession.
Although this may seem extreme, such values are seen on many occasions.
The reason for such behaviour is that the starting values of eccentricity
and inclination may be highly favourable, which makes for large values of
$e_{\rm max}$.

The relevant equations of motion are integrated by the fourth-order
Runge-Kutta method, taking special care with the time-step choice for large
eccentricities.
After each integration interval, which usually consists of many small
steps, the new semi-major axis is determined by
$a = {\bf h}^2 /(1 - e^2) (m_1 + m_2)$, whereupon the relative quantities
${\bf r}$ and ${\bf v}$ are determined.
This enables the KS elements ${\bf u}$ and ${\bf u}'$ to be derived from
standard transformations.
The latter procedure is required for the purpose of a consistent
re-activation of the inner binary at termination.

Apart from terminating the averaging treatment because of large external
perturbations or violation of the stability criterion, we also switch to
direct KS integration of the inner binary if the circularization time,
$t_{\rm circ}$, falls below the standard value $2 \times 10^9~yr$
(cf.\ Mardling \& Aarseth 2001).
The subsequent evolution is then subject to the tidal circularization
process by itself, whereas the oblateness and relativistic precession is
not taken into account.
Provided the value of $t_{\rm circ}$ is sufficiently small, the ongoing
Kozai cycle is included via the perturbation.
Hence the end result is influenced by the ratio $t_{\rm Kozai}/t_{\rm circ}$,
where the former increases through orbital shrinkage and the latter has
a steep dependence on $R_p$ as well as the stellar radii which tend to grow.
A typical outcome is for the hierarchy to be disrupted after some time,
whereupon the binary undergoes normal circularization.
However, the possible effect of the neglected apsidal motion during the
circularizing stage still needs to be clarified.
The main reason for adopting the averaging treatment is that the outer
component may induce a large eccentricity which is followed by a stage of
significant orbital shrinkage by angular momentum conservation.
Consequently, such interactions may provide a mechanism for producing close
binaries.

\section{Case Studies}

The cluster models contain a wealth of data which may be used to
illustrate many aspects of interest.
In the following, we concentrate on describing some characteristic systems 
which exhibit period shrinkage.
To set the scene, we first present a snapshot of the short-period
hierarchical binaries at a typical epoch.
Thus near the cluster half-life of $t \simeq 2~Gyr$, the remaining
population is given by $N_s = 5020, \,N_b = 1200$.
At this stage there are 14 stable hierarchies with a pronounced central
concentration; i.e.\ five are inside the core radius and 13 inside the
half-mass radius.
Five of these systems contain an inner binary with short period, as
illustrated in Table~1.
Here the initial and final inner period is given in Columns 1 and 2 and
$T_{\rm out}$ is the current outer period.
All these binaries experienced the orbital shrinkage as a result of the
averaging procedure, with three in the final stages of circularization.

\begin{table}
\caption{Hierarchical binaries}
\begin{center}
\begin{tabular}{rlllll}
\tableline
     $ T_0$ & $T_f$ & $T_{\rm out}$ & $e_0$ & $e_f$ & remarks~~~~~~\\
\tableline
6 & 2.3 & $1 \times 10^4$ & 0.59 & 0.15 &circularizing \\
200 & 1.6 & $6 \times 10^4$ & 0.38 & 0.01 &circularizing \\
6000 & 1.3 & $8 \times 10^6$ & 0.84 & 0.11 &circularizing \\
2000 & 1.1 & $7 \times 10^4$ & 0.83 & 0.00 \\
290 & 0.5 & $3 \times 10^4$ & 0.65 & 0.00 \\
\tableline\tableline
\end{tabular}
\end{center}
\end{table}

In order to highlight the process of period reduction, we describe briefly
some evolutionary sequences.
The first case study consists of a parabolic binary-binary collision
resulting in exchange with new semi-major axis of $5~AU$.
A hierarchy is formed with periods
$T_{\rm in} \simeq 2000, \,T_{\rm out} \simeq 10^5~d$ and eccentricity
$e_{\rm out} \simeq 0.7$, and hence is quite stable.
This leads to an induced inner eccentricity $e_f = 0.982$, which triggers
chaotic tides until normal tidal dissipation takes over when $e_f = 0.82$.
During the next $12~Myr$ until complete circularization, the period is
reduced from $30~d$ to $8~d$.
The ultimate fate is coalescence by common envelope evolution after an
interval of Roche-lobe mass transfer lasting $\simeq 10^8~yr$.

Unstable multiple systems may also produce interesting outcomes of period
reduction, as occurred in another example.
Here the interaction of a single particle and a binary of period
$T_0 \simeq 7 \times 10^4~d$ resulted in exchange with
$T \simeq 6 \times 10^4~d$ and high eccentricity, $e = 0.9997$.
Chaotic tidal evolution reduced the period to $T_f = 3.5~d$, whereupon
normal circularization ended with a final period $T_f = 0.6~d$ after an
interval of $2~Myr$.
This close binary was retained by the cluster over the next $3~Gyr$ until
it escaped.

Some hierarchical systems may be involved in complicated evolution paths,
as in the following sequence.
A primordial binary started with the initial elements
$T_0 = 1977~d, \, e_0 = 0.83, \, m_1 = 1.4, \, m_2 = 0.7~m_{\odot}$.
A coasting interval of $1.0~Gyr$ hardly produced any change (i.e.\
$T = 1982, \, e = 0.84$) until a binary-binary collision created a stable
triple with $T_{\rm out} \simeq 9 \times 10^4~d$ and $e_{\rm out} = 0.66$.
The induced eccentricity reached $e = 0.993$, whereupon circularization
was initiated leading to $T_f = 1.1~d$ after $12~Myr$.
Much later, at epoch $t \simeq 2.4~Gyr$ Roche mass transfer took place,
with the inner binary undergoing coalescence without mass loss.
Another binary-binary collision occurred, with new hierarchical parameters
$T_{\rm in} \simeq 4000, \,T_{\rm out} \simeq 5 \times 10^5~d$ and
$e_{\rm out} = 0.51$.
Again chaotic and normal tides started from a high eccentricity of 0.989
and reduced the inner period to $32~d$.
This soon gave rise to a first Roche stage involving a giant and white
dwarf, followed by a second Roche stage of the AGB and white dwarf at epoch
$t \simeq 3.2~Gyr$.
The ensuing common envelope evolution resulted in another white dwarfs with
$a_f \simeq 1~r_{\odot}$ and period $0.1~d$.
Finally, at epoch $t \simeq 4.3~Gyr$ the binary decayed by gravitational
radiation braking to a period $T_f \simeq 0.001~d$ and ultimate coalescence.
Hence the initial binary was involved in two separate episodes of
binary-binary collisions as well as two circularization stages during an
interval exceeding $4~Gyr$, compared to a total disruption time of nearly
$6~Gyr$.
This example also demonstrates the intricate relation between the dynamics
and stellar evolution which takes place in a star cluster.

\section{Conclusions}

The first part of this paper deals with the formation of hierarchical
systems in star cluster models containing a significant proportion of
primordial binaries.
Provided the orbital characteristics satisfy the stability criterion given
by Eq.~(1) and the binding energy of the outer component is fairly hard,
a triple is represented by the corresponding two-body motion.
We find that typically at least ten such systems are present after the
first $Gyr$ in simulations with $N_s = 8000$ single and $N_b = 2000$ binary
stars.
The stable triples form as the result of binary-binary collisions, often
resulting in hard outer energies and are therefore quite robust.
Moreover, the relatively small formation rate is compensated by long
life-times.
On the other hand, most quadruples originate via classical three- and
multi-body processes and are therefore less well bound initially.
Given the presence of such systems, further interactions may produce more
complex structures like quintuplets and sextuplets, especially during late
stages of the cluster life-time.
Thus we may invoke the concept of dynamical molecules when describing the
different outcomes of interactions (cf.\ Aarseth 1988).
Consequently, the internal degrees of freedom of high-order systems
represent latent energy which becomes available for heating the cluster on
disruption.
We also emphasize that hierarchical systems have large cross sections and
tend to be centrally concentrated.
Since such configurations originate by a combination of processes, their
formation is best studied {\it in situ}, rather than by scattering
experiments which have so far only reached the case $B + B$ in complexity.

In the second part, we outline a scheme for modelling the eccentricity
change of the inner binary due to the outer component moving in an inclined
orbit.
Thus if the inclination is favourable, the induced eccentricity may attain
a large value, with a corresponding small pericentre separation, and
thereby initiate tidal dissipation.
In some cases this leads to significant orbital shrinkage at the end of the
circularization and is therefore a mechanism for producing close binaries.
Our development is based on calculating the effect of the outer body on
the inner eccentricity and angular momentum vectors, using a quadrupole
expansion.
In addition, we include contributions from the internal tidal dissipation
and apsidal motion due to oblateness, as well as relativistic precession.
So far the averaging treatment appears to be promising but should still be
considered experimental.

\acknowledgments

We thank Peter Eggleton and Chris Tout for valuable advice
on all aspects of stellar evolution and Seppo Mikkola for his pioneering
contributions to the regularization methods which are so vital for this work.

\end{document}